\documentclass[letterpaper, 10 pt, conference]{ieeeconf}

\IEEEoverridecommandlockouts

\usepackage{balance}

\usepackage{multirow}

\usepackage{amsmath,mathtools}  %
\usepackage{amsfonts,amssymb}
\usepackage{bbm}

\usepackage{xspace}    %

\usepackage{graphicx}
\graphicspath{{figures/}}

\usepackage{subcaption}
\usepackage[font=footnotesize]{caption}

\usepackage{hyperref}
\hypersetup{colorlinks=true, linkcolor= blue, allcolors=blue, citecolor = blue, filecolor=black, urlcolor=blue}

\usepackage{todonotes}
\usepackage{comment}

\usepackage{float}%

\usepackage{tikz}
\usetikzlibrary{arrows,positioning}   %

\usepackage{pgfplots}\pgfplotsset{compat=1.9}
\usepackage{grffile}
\pgfplotsset{compat=newest}  %
\usetikzlibrary{plotmarks}
\usetikzlibrary{arrows.meta}
\usepgfplotslibrary{patchplots}
\usepgfplotslibrary{groupplots}

\pgfplotsset{compat=1.15}
\usepackage{mathrsfs}
\usetikzlibrary{arrows}

\usepackage{tabularx}

\usepackage[ruled]{algorithm}
\usepackage[noend]{algpseudocode}

\makeatletter
\newcommand{\multiline}[1]{%
  \begin{tabularx}{\dimexpr\linewidth-\ALG@thistlm}[t]{@{}X@{}}
    #1
  \end{tabularx}
}
\makeatother

\usepackage[per-mode=symbol]{siunitx}

\usepackage{cite}  %
\usepackage{array}  %
\usepackage{booktabs}           %

\newcounter{remark}
{\par\endtrivlist\unskip}

\newtheorem{definition}{Definition}

\newcounter{problem}
\newenvironment{problem}{%
\par\vspace{3pt}\noindent\refstepcounter{problem}\textbf{Problem~\theproblem:}}%
{\par\endtrivlist\unskip}

\usepackage[extramath,probability,reviewmark]{truonglatexdefs}

\newcommand{\CAV}[1]{CAV\textendash \ensuremath{#1}\xspace}
\newcommand{\HDV}[1]{HDV\textendash\ensuremath{#1}\xspace}

\usepackage{stfloats}%

\title{\LARGE \bf Coordination for Connected Automated Vehicles at Merging Roadways in Mixed Traffic Environment}
\author{Viet-Anh Le$^{1,2}$, {\IEEEmembership{Student Member, IEEE}}, Hao M. Wang$^{3}$, {\IEEEmembership{Student Member, IEEE}}, \\G\'abor Orosz$^{3,4}$, {\IEEEmembership{Senior Member, IEEE}}, and Andreas A. Malikopoulos$^{5}$, {\IEEEmembership{Senior Member, IEEE}}
\thanks{This work was supported by NSF under Grants CNS-2149520 and CMMI-2219761.}
\thanks{$^{1}$Department of Mechanical Engineering, University of Delaware, Newark, DE 19716 USA.}
\thanks{$^{2}$Systems Engineering, Cornell University, Ithaca, NY 14850 USA.}
\thanks{$^{3}$Department of Mechanical Engineering, University of Michigan, Ann Arbor, MI 48109 USA.}
\thanks{$^{4}$Department of Civil and Environmental Engineering, University of Michigan, Ann Arbor, MI 48109 USA.}
\thanks{$^{5}$School of Civil and Environmental Engineering, Cornell University, Ithaca, NY 14850 USA.}
\thanks{Emails: {\tt\small vl299@cornell.edu}, {\tt\small haowangm@umich.edu}, {\tt\small orosz@umich.edu}, and {\tt\small amaliko@cornell.edu}.}
}

\begin{document}

\maketitle
\thispagestyle{empty}
\pagestyle{empty}

\begin{abstract}
In this paper, we present an optimal control framework to address motion coordination of connected automated vehicles (CAVs) in the presence of human-driven vehicles (HDVs) in merging scenarios. 
Our framework combines an unconstrained trajectory solution of a low-level energy-optimal control problem with an upper-level optimization problem that yields the minimum travel time for CAVs.
We predict the future trajectories of the HDVs using Newell's car-following model. 
To handle potential deviations of HDVs' actual behavior from the predicted one, we design a safety filter for CAVs based on control barrier functions.
The effectiveness of the proposed control framework is demonstrated via simulations with heterogeneous human driving behaviors.
\end{abstract}

\section{Introduction}\label{sec:intro}

Emerging vehicular technologies in automation and communication have generated new opportunities to enhance traffic safety and efficiency \cite{ersal_connected_2020}.
Prior research efforts have shown that in environments consisting solely of connected automated vehicles (CAVs), the traffic throughput and energy consumption can be significantly improved under different vehicle coordination strategies such as optimal control \cite{Malikopoulos2020,JinGabor2017,chalaki2020experimental}, model predictive control \cite{katriniok2022fully}, control barrier functions \cite{ChalakiCBF2022}, and learning-based control \cite{chalaki2020ICCA,chalaki2020hysteretic}.
It is expected that CAVs will gradually penetrate the market and interact with human-driven vehicles (HDVs) over the next several years. 
The presence of HDVs, however, complicates vehicle coordination, given the existing limitations on predicting precise driving behavior \cite{wang_social_2022}.

Addressing coordination of CAVs in a mixed traffic environment, where CAVs and HDVs co-exist, has attracted considerable attention recently. 
Some examples include game theory combined with reinforcement learning \cite{Yildiz2019}, dynamic programming \cite{Papageorigiou2020TRC}, statistical modeling of human uncertainties \cite{Tomlin2019}, and reachability analysis \cite{Althoff2021,Sanghoon2023ITSC}. Efforts on experimental validation with scaled robotic cars have been reported in \cite{chalaki2021CSM}.
In mixed autonomy, traffic bottlenecks such as merging roadways and intersections pose significant challenges to the decision-making and control of CAVs. 
A centralized algorithm for socially compliant navigation at an intersection given the social preferences of the vehicles was presented in \cite{buckman_sharing_2019}. 
A hierarchical on-ramp merging control strategy for CAVs was presented in \cite{liu2023safety}, where an upper-level planner computes an expected merging position and a low-level trajectory controller guarantees constraints under uncertainties of HDVs.
A vehicle coordination problem for a signal-free intersection using vehicle platooning was addressed in \cite{faris_optimization-based_2022}. 

In this paper, we extend the framework developed in \cite{Malikopoulos2020,chalaki2020experimental} for CAV coordination to a mixed-traffic merging scenario.
We present an optimal control formulation to derive the trajectories for the CAVs, in which the unconstrained trajectory solution of a low-level energy-optimal control problem is embedded in an upper-level optimization problem aimed at finding the minimum travel time for the CAV. 
We employ Newell's car-following model \cite{newell_simplified_2002} to predict the HDVs' future trajectories, which are used to formulate no-conflict constraints in the optimal control problem. 
The constraints help each CAV avoid conflicts with the vehicles traveling on the same road and those traveling on the other merging road. 
Since the predicted trajectories may deviate from those arising from the human drivers' actions, we design a safety filter for the CAVs based on control barrier functions \cite{CBF_2023,Tamas2023CDC}.
The safety filter modifies the original optimal control input of the CAV in a minimally invasive fashion so that conflict-free maneuvers are ensured. 
We demonstrate the efficacy of the proposed framework by simulations given different levels of CAV penetration and traffic volumes.

The remainder of the paper proceeds as follows.
In Section~\ref{sec:problem}, we formulate the coordination problem of CAVs in a mixed-traffic merging scenario. 
In Section~\ref{sec:optimal_control}, we present the optimal control problem formulation combined with a method for predicting the future trajectories of HDVs, and a control barrier function-based safety filter.
The simulation results are provided in Section~\ref{sec:sim}. 
Finally, we draw concluding remarks in Section~\ref{sec:conc}.

\section{Problem Formulation} \label{sec:problem}

\begin{figure}[!t]
  \begin{center}
  \includegraphics[scale=0.35]{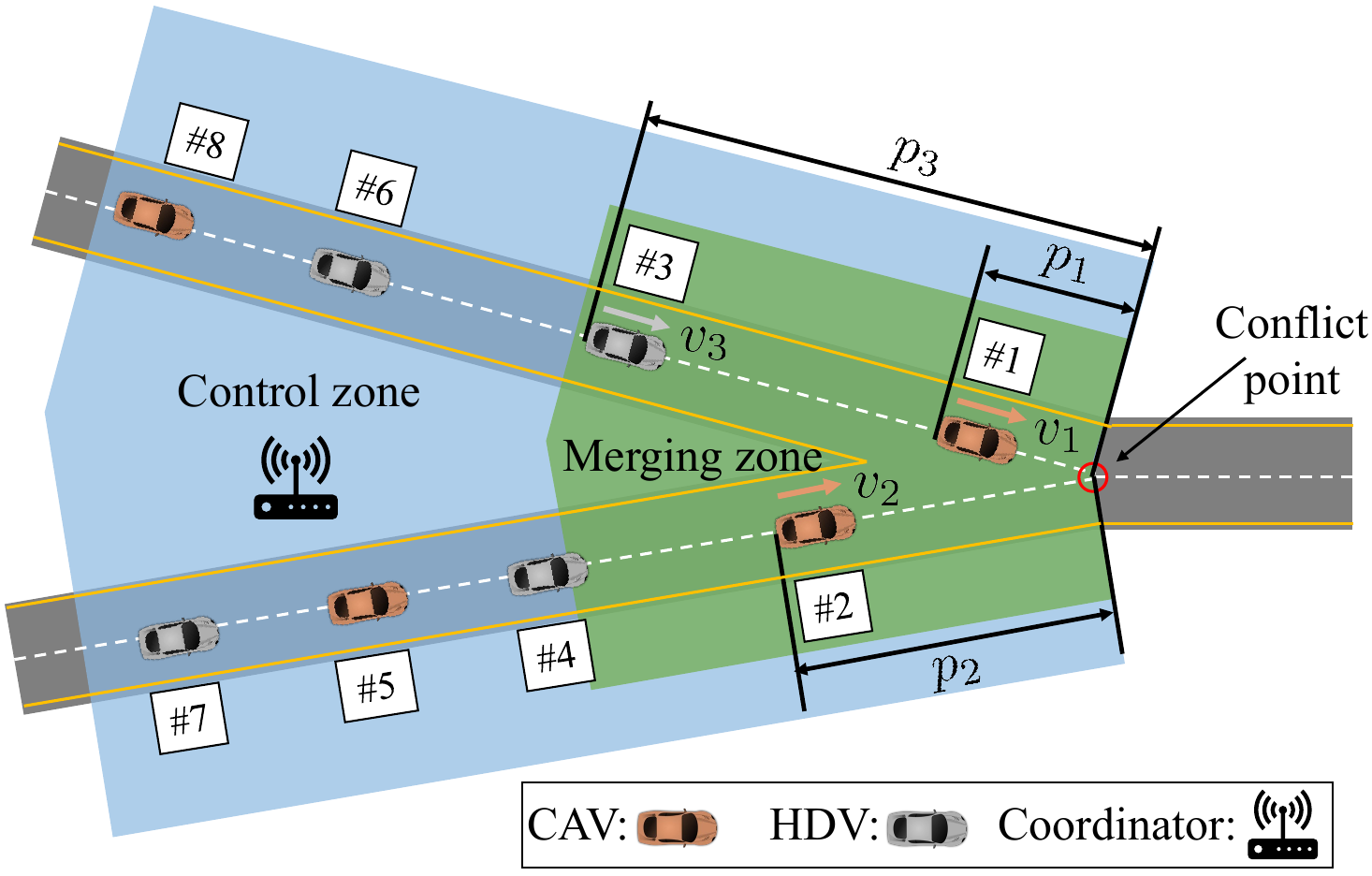}
  \end{center}
\caption{Merging scenario with two merging roadways intersecting at a conflict point. The blue area represents the control zone. In the merging zone (green area), virtual projection is utilized as detailed in Section~\ref{sec:hdv}.} 
\label{fig:scenario}
\vspace{-5mm}
\end{figure}

We consider the problem of coordinating multiple CAVs, co-existing with HDVs, in a scenario where two merging roadways intersect at a conflict point; see Fig.~\ref{fig:scenario}. 
We define a control zone inside which a coordinator is able to monitor the motion of all vehicles (including CAVs and HDVs). 
Communication errors or delays are neglected. 
The analysis of such effects is left for future research.
Next, we provide the following necessary definitions for our exposition.

\begin{definition}
\label{def:set_vehicle}
Let ${\LLL(t) = \{1,\ldots,L(t)\}}$, ${t\in\mathbb{R}_{\ge0}}$, be the set of vehicles traveling inside the control zone at time~$t$, where ${{L}(t)\in\mathbb{N}}$ is the total number of vehicles. 
Let ${\AAA(t) \subset \LLL(t)}$ and ${\HHH(t) \subset \LLL(t)}$ be the sets of CAVs and HDVs, respectively.
For example, considering the scenario in Fig.~\ref{fig:scenario}, ${\LLL(t) = \{ 1, \dots, 8\}}$, ${\AAA(t) = \{ 1, 2, 5, 8\}}$, and ${\HHH(t) = \{ 3, 4, 6, 7\}}$.  
Note that the indices of the vehicles are determined by the order in which they enter the control zone. 
\end{definition}
\begin{definition}
For a vehicle ${i\in\LLL(t)}$, let ${\SSS_i(t) \subset \LLL (t)}$ and ${\NNN_i(t) \subset \LLL (t)}$, ${t\in\mathbb{R}_{\ge0}}$, be the sets of vehicles inside the control zone traveling on the same road as vehicle~$i$ and on the neighboring road, respectively.
\end{definition}

We consider that the dynamics of vehicle~${i \in \LLL (t)}$ are described by a double integrator model:
\begin{equation}\label{eq:model2}
\begin{split}
\dot{p}_{i}(t) &= v_{i}(t), 
\\
\dot{v}_{i}(t) &= u_{i}(t), 
\end{split}    
\end{equation}
where ${p_{i}\in\mathcal{P}}$, ${v_{i}\in\mathcal{V}}$, and
${u_{i}\in\mathcal{U}}$ denote the longitudinal position of the rear bumper, speed, and control input (acceleration) of the vehicle, respectively. 
The sets ${\mathcal{P}, \mathcal{V},}$ and $\mathcal{U}$ are compact subsets of $\RR$. 
Note that we set ${p_{i} = 0}$ at the conflict point; see Fig.~\ref{fig:scenario}.
The control input is bounded by 
\begin{equation}\label{eq:uconstraint}
u_{\min} \leq u_{i}(t)\leq u_{\max},\quad \forall i \in \LLL(t),
\end{equation}
where ${u_{\min}<0}$ and ${u_{\max}>0}$ are the minimum and maximum control inputs given by the physical acceleration and braking limits of the vehicles or imposed by driver/passenger comfort.
Next, we provide the speed limits of the CAVs, 
\begin{equation}\label{eq:vconstraint} 
0 \leq v_i(t) \leq v_{\max},\quad \forall i \in \AAA (t), 
\end{equation}
where ${v_{\max}>0}$ is the maximum allowable speed. 
We do not impose a maximum speed limit for HDVs since human drivers may violate it, but we still assume that
\begin{equation}
0 \leq v_i(t),\quad \forall i \in \HHH (t),
\end{equation}
which means the HDVs do not go backward.

To avoid conflicts between vehicles, we impose two types of constraints: (i) between vehicles traveling on neighboring roads (to avoid that they meet at the conflict point) and (ii) between vehicles traveling on the same road (to avoid rear-end collisions).
To prevent a potential conflict between \CAV{i} and a vehicle ${k \in \NNN_i(t)}$ traveling on the neighboring road, we require a minimum time gap ${t_{\min}>0}$ between the time instants ${t_i^\mathrm{f}}$ and $t_k^\mathrm{f}$ when the \CAV{i} and vehicle~$k$ cross the conflict point, respectively, i.e.,
\begin{equation}\label{eq:lateral_constraint}
\abs{t_i^\mathrm{f} - t_k^\mathrm{f}} \geq t_{\min}. 
\end{equation}
To prevent rear-end collision between \CAV{i} and its immediate preceding vehicle~$k$ traveling on the same road, i.e., ${k = \max \, \{ j \in \SSS_i(t)\; | \; j < i \}}$, we impose the constraint:
\begin{equation}\label{eq:rearend_constraint}
 p_{k}(t) - p_{i}(t) \geq  d_{\min} + t^\mathrm{h}_{\min} v_{i}(t),
\end{equation}
where ${d_{\min}>0}$ and ${t^\mathrm{h}_{\min}>0}$ are the minimum standstill distance and the minimum time headway.
Note that we use the distance between the vehicles' rear bumpers, and the vehicle length is included by choosing sufficiently large ${d_{\min}}$.

\section{Optimal Control Formulation}\label{sec:optimal_control}

In this section, we present an optimal control framework to coordinate the CAVs in mixed traffic, extending the one developed for 100\% CAV penetration in \cite{Malikopoulos2020,chalaki2020experimental}.
For each CAV, we use the unconstrained trajectory solution of a low-level energy-minimal optimal control problem to formulate an upper-level optimization problem that finds the minimum time for crossing the conflict point while satisfying all state, control, and safety constraints.

\subsection{Optimal Control Problems}

We formulate the low-level optimization by considering that the solution to the upper-level optimization problem is known, i.e., that the minimum time $t_i^\mathrm{f}$ which satisfies all constraints is given. 
Then, the low-level optimal control problem aims at finding the control input (acceleration) for each CAV by solving the following optimization problem.
\begin{problem}\label{prb:energy-optimal-1}
(\textbf{Low-level energy-optimal control})
Let $t_{i}^0$ and $t_{i}^\mathrm{f}$ be the times that \CAV{i} enters and exits the control zone, respectively. 
Then,  \CAV{i} solves the following optimal control problem at $t_{i}^0$: 
\begin{equation}\label{eq:energy_cost}
\begin{split}
&\minimize_{u_i(t)\in \UUU} \quad \frac{1}{2} \int_{t^{0}_{i}}^{t^\mathrm{f}_{i}} u^2_i(t)~\mathrm{d}t,
\\ 
& \subjectto 
\\
& \quad \eqref{eq:model2}, \eqref{eq:uconstraint},\eqref{eq:vconstraint},
\\
& \quad \eqref{eq:rearend_constraint},\; k = \max \, \{ j \in \SSS_i(t_i^0)\; | \; j < i \}, 
\\
& \text{given:} 
\\
& \quad p_i (t_i^0) = p^0, \,\, v_i (t_i^0) = v_i^0, 
 \,\, p_i (t_i^\mathrm{f}) = 0, \,\, u_i (t_i^\mathrm{f}) = 0,
\end{split}
\end{equation}
where $p^0$ is the position of the entry point of the control zone.
The boundary conditions in \eqref{eq:energy_cost} are set at the entry and exit of the control zone. 
\end{problem}

Note that \eqref{eq:lateral_constraint} is not included in the low-level problem since the time $t_{i}^\mathrm{f}$, derived through the upper-level problem discussed below, satisfies \eqref{eq:lateral_constraint}.
Also, note that in \eqref{eq:energy_cost} we minimize the $L^2$ norm of control input $u_i(t)$ in ${t \in [t_{i}^0,\ t_{i}^\mathrm{f}]}$ while  the energy consumption per unit mass is given by
\begin{equation}
\label{eq:energy_consumption}
\varepsilon = \int_{t^{0}_{i}}^{t^\mathrm{f}_{i}} v_i(t) g(u_i(t))~\mathrm{d}t,
\end{equation}
where ${g(x)=\max\{0,x\}}$.
However, it was shown in \cite{Minghao2023ITS} that the energy consumption \eqref{eq:energy_consumption} can be upper and lower bounded by monotonic functions of the cost function in \eqref{eq:energy_cost}. %
Therefore, minimizing the $L^2$ norm of the control input still benefits energy efficiency.

The closed-form solution of Problem~\ref{prb:energy-optimal-1} can be derived using the Hamiltonian analysis \cite{Malikopoulos2020}.
In case none of the state and control constraints are active, the unconstrained Hamiltonian is formulated as
\begin{equation}\label{c1}
\begin{aligned}
\!\! H_i(t,p_i(t),v_i(t),u_i(t)) = \frac{1}{2}u_i^2(t) + \lambda_i^p v_i (t) + \lambda_i^v u_i(t),
\end{aligned}
\end{equation}
where \(\lambda_i^p\) and \(\lambda_i^v\) are co-states corresponding to position and speed, respectively.
The Euler-Lagrange equations of optimality are given by
\begin{align}
\dot{\lambda}_i^p & =-\frac{\partial H_i}{\partial p_i}=0, \label{euler:optimallambdav} 
\\ 
\dot{\lambda}_i^v &= -\frac{\partial H_i}{\partial v_i}= -\lambda_i^p, \label{euler:optimallambdap}
\\
\frac{\partial H_i}{\partial u_i} &= u_i+\lambda_i^v=0 \label{euler:optimalU}.
\end{align}
Since the speed of \CAV{i} is not specified at the travel time $t_i^{\mathrm{f}}$, we have the boundary condition
\begin{equation}\label{eq:lambdaV-sb}
    \lambda_i^v(t_i^{\mathrm{f}}) = 0.
\end{equation}
Applying the Euler-Lagrange optimality conditions \eqref{euler:optimallambdav}-\eqref{euler:optimalU} to the Hamiltonian \eqref{c1}, yields the optimal control law
\begin{equation}\label{eq:optimalControl}
u_i^\ast(t) = -{\lambda_i^v}^\ast = 6 a_it + 2b_i,
\end{equation}
where $a_i$ and $b_i$ are constants of integration. 
Therefore, the unconstrained solution takes the form
\begin{equation}\label{eq:optimalTrajectory}
\begin{split}
v^\ast_i(t) &= 3 a_i t^2 + 2 b_i t + c_i, 
\\
p^\ast_i(t) &= a_i t^3 + b_i t^2 + c_i t + d_i,
\end{split}
\end{equation} 
where ${c_i, d_i \in \RR}$ are constants of integration.
Note that substituting \eqref{eq:lambdaV-sb} into \eqref{euler:optimalU} at $t_i^{\mathrm{f}}$ yields the terminal condition ${u_i(t_i^{\mathrm{f}}) = 0}$.
Given the boundary conditions in \eqref{eq:energy_cost}, 
if $t_i^\mathrm{f}$ is known, the constants of integration can be found as
\begin{equation}
\label{eq:phi_i}
\begin{bmatrix}
a_i 
\\
b_i 
\\
c_i 
\\
d_i
\end{bmatrix}
= 
\begin{bmatrix}
(t_i^0)^3 & (t_i^0)^2 & t_i^0 & 1 
\\
3(t_i^0)^2 & 2t_i^0 & 1 & 0 
\\
(t_i^\mathrm{f})^3 & (t_i^\mathrm{f})^2 & t_i^\mathrm{f} & 1 
\\
6t_i^\mathrm{f} & 2 & 0 & 0 
\end{bmatrix}^{-1}
\begin{bmatrix}
p^0 
\\
v_i^0 
\\
0 
\\
0
\end{bmatrix},
\end{equation}
in case the matrix above is not singular.

Next, we formulate the upper-level optimal control problem to minimize the travel time and guarantee all the constraints for the energy-optimal trajectory \eqref{eq:optimalTrajectory}; see \cite{malikopoulos2018decentralized}.
\begin{problem} 
(\textbf{Upper-level minimal-time planning})
Once entering the control zone, \CAV{i} solves the following optimal control problem at $t_i^0$ \label{prb:optimal_MZ}
\begin{align}
\begin{split}
    \label{eq:tif_1}
    &\minimize_{t_i^\mathrm{f} \in \mathcal{T}_i(t_i^0)} \quad t_i^\mathrm{f} 
    \\
    & \subjectto 
    \\
    & \quad \eqref{eq:uconstraint}, \eqref{eq:vconstraint}, 
    \\
    & \quad \eqref{eq:lateral_constraint},\; \forall \, k \in \NNN_i(t_i^0), 
    \\
    & \quad \eqref{eq:rearend_constraint},\; k = \max \, \{ j \in \SSS_i(t_i^0)\; | \; j < i \}, 
    \\
    & \quad \eqref{eq:optimalTrajectory}, 
    \\
    & \text{given:} 
    \\
    & \quad p_i (t_i^0) = p^0, \,\, v_i (t_i^0) = v_i^0, 
   \,\, p_i (t_i^\mathrm{f}) = 0, \,\, u_i (t_i^\mathrm{f}) = 0. 
\end{split}
\end{align}
Here, the compact set ${\mathcal{T}_i(t_i^0)=[\underline{t}_i^\mathrm{f}, \overline{t}_i^\mathrm{f}]}$ represents the feasible range of travel time under the state and input constraints of \CAV{i} computed at $t_i^0$. 
\end{problem}

The computation steps to solve Problem~\ref{prb:optimal_MZ} numerically are summarized next; for more details, see \cite{chalaki2020experimental}.
First, we initialize ${t_i^\mathrm{f} = \underline{t}_i^\mathrm{f}}$, and compute the parameters ${[a_i,b_i,c_i,d_i]}$ using \eqref{eq:phi_i}.
We evaluate all the state, control, and no-conflict constraints. If none of the constraints is violated, we return the solution; otherwise, $t_i^\mathrm{f}$ is increased by a step size. 
The procedure is repeated until the solution satisfies all the constraints.
By solving Problem~\ref{prb:optimal_MZ}, the optimal exit time  $t_i^\mathrm{f}$ along with the optimal trajectory \eqref{eq:optimalTrajectory} and control law \eqref{eq:optimalControl} are obtained for \CAV{i} for ${t\in [t_i^0, t_i^\mathrm{f}]}$.

If a feasible solution to Problem~\ref{prb:optimal_MZ} exists, then the solution is a cubic polynomial of the form \eqref{eq:optimalTrajectory}, guaranteeing that none of the constraints are activated.
If the solution of Problem~\ref{prb:optimal_MZ} does not exist,
one may derive an alternative trajectory numerically by piecing together the constrained and unconstrained arcs \cite{malikopoulos2018decentralized}.

\subsection{Human Drivers' Trajectory Prediction}\label{sec:hdv}

To solve Problem~\ref{prb:optimal_MZ}, all vehicles' trajectories and exit times having potential conflicts with \CAV{i} must be available.
When \CAV{i} enters the control zone, the trajectories and exit times of other CAVs traveling in the control zone can be obtained from the coordinator via wireless communication.
The states of HDVs are assumed to be available, but their future trajectories are not known.

Next, we present an approach to predict the future trajectories and exit times of the HDVs traveling in the control zone based on Newell's car-following model \cite{newell_simplified_2002}.
This model simply considers that due to traffic wave propagation, the trajectory of a vehicle is a shifted copy in time and space of its predecessor.
Specifically, the position of each \HDV{k}, ${k \in \HHH (t)}$, is predicted from the position of its preceding vehicle ${j = \max \, \{ l \in \SSS_i(t)\; | \; l < i \}}$ as
\begin{equation}\label{eq:newell}
p_{k}(t) = p_{j} \big( t - \tau_{k} \big) - w\, \tau_{k},
\end{equation}
where ${\tau_{k}>0}$ is the time shift of \HDV{k}, and ${w > 0 }$ is the speed of the backward propagating traffic waves, which is considered to be constant \cite{molnar_board_2022}.
Here we use ${w = \SI{5}{m/s}}$.

Since ${v_{j} (t)\geq 0}$ and ${w > 0}$, ${p_{j} \left( t - \tau_{k} \right) - w \tau_{k}}$ is a strictly decreasing function of $\tau_{k}$. 
Thus, there exists a unique value of $\tau_{k}$ such that \eqref{eq:newell} is satisfied for any $t$. 
That is, when \CAV{i} enters the conflict zone at ${t=t_i^0}$, it can solve \eqref{eq:newell} for $\tau_k$ for each \HDV{k} based on the positions of the vehicles (obtained from the coordinator).
If there is no preceding vehicle in front of \HDV{k}, it is assumed to maintain its current speed, making its position an affine function of time.

\begin{figure}[!t]
  \begin{center}
  \includegraphics[scale=0.38]{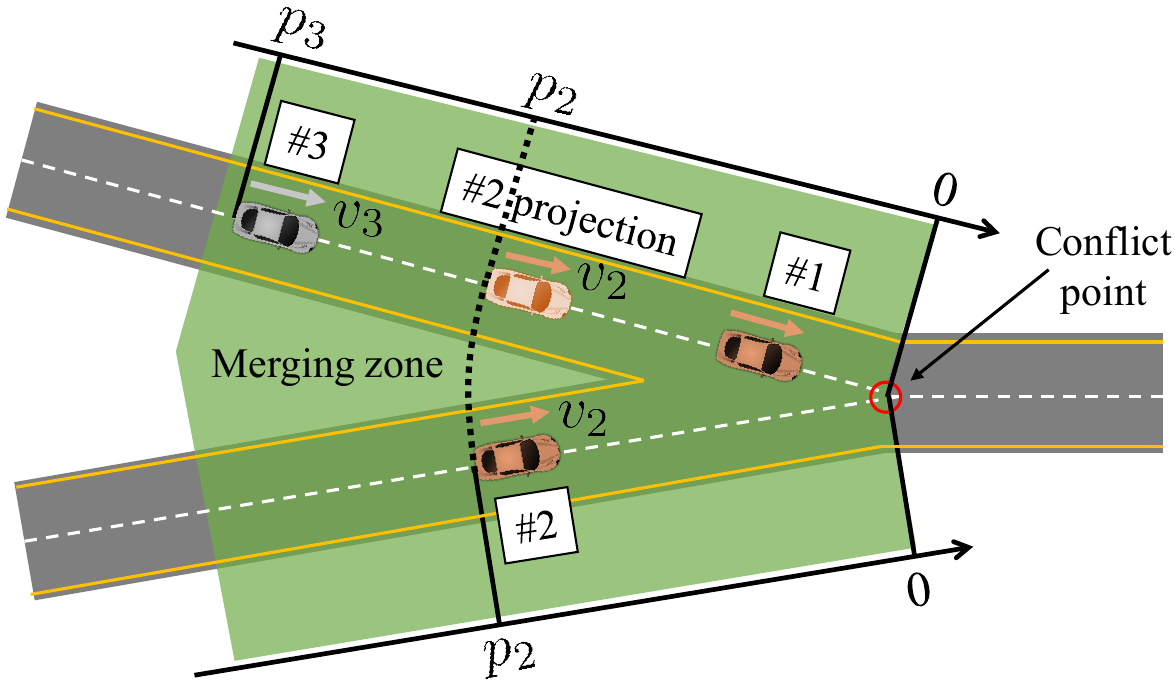}
  \end{center}
\caption{An example of virtual projection in the merging zone, where the \CAV{2} is projected from the perspective of \HDV{3}. } 
\label{fig:projection}
\vspace{-5mm}
\end{figure}

In the proximity of the conflict point, defined as the merging zone in Fig.~\ref{fig:scenario}, the HDVs interact with vehicles on the neighboring road and control their longitudinal motion accordingly. 
This is achieved by the virtual projection of vehicles traveling on the neighboring road, an example shown in Fig.~\ref{fig:projection}. 
Here, from the perspective of \HDV{3}, the projected \CAV{2} is considered as the preceding vehicle instead of \CAV{1}.
Similar generalized car-following models for merging and lane-changing scenarios have been reported in \cite{VirtualPlatoon,guo2019improved}.
 
Using the proposed prediction model, the position of each HDV in the control zone can be represented either by a cubic polynomial or by an affine function of time. 
The trajectory prediction is then used for estimating the HDVs' exit times to impose the no-conflict \eqref{eq:lateral_constraint} and rear-end safety constraints \eqref{eq:rearend_constraint}.
We use $t_k^\mathrm{f}$ and ${[a_k,b_k,c_k,d_k]}$, ${k \in \HHH (t)}$, to denote the predicted exit time and the constants parameterizing the predicted trajectory of \HDV{k}.  
Substituting the solution \eqref{eq:optimalTrajectory} into \eqref{eq:newell} (for both $p_k$ and $p_j$), the trajectory parameters of \HDV{k} can be obtained from those of the preceding vehicle~$j$ according to
\begin{equation}\label{eq:newell-params}
\begin{split}
a_k &= a_j, 
\\ 
b_k &= b_j - 3a_j \tau_k, 
\\
c_k &= c_j + 3a_j\tau_k^2 - 2b_j \tau_k, 
\\
d_k &= d_j - a_j \tau_k^3 + b_j \tau_k^2 - c_j \tau_k - w \tau_k,
\end{split}
\end{equation}
and $t_k^\mathrm{f}$ can be found by solving 
\begin{equation}
a_k (t_k^\mathrm{f})^3 + b_k (t_k^\mathrm{f})^2 + c_k t_k^\mathrm{f} + d_k = 0.
\end{equation}
Then $t_k^\mathrm{f}$ and ${[a_k,b_k,c_k,d_k]}$
can be integrated into the constraints \eqref{eq:lateral_constraint} and \eqref{eq:rearend_constraint} between 
\CAV{i} and \HDV{k}.

\subsection{Safety Filter using Control Barrier Functions}
\label{subsec:cbf}

Using Newell’s car-following model, we can predict the HDVs’ future trajectories and then integrate them into Problem~\ref{prb:optimal_MZ} to derive the trajectory and control law for CAVs in the control zone. 
However, in reality, HDVs may behave differently from the predicted model.
Under this discrepancy, the derived optimal control law for CAVs may not always ensure conflict-free maneuvers.
To address this issue, in this subsection, we utilize control barrier functions \cite{CBF_2023} to design a safety filter that modifies the original optimal control input to a safe control action once unsafe situations are realized. 
Thus, conflict-free maneuvers are guaranteed for the CAVs.

\begin{figure}[!t]
  \begin{center}
  \includegraphics[scale=0.4]{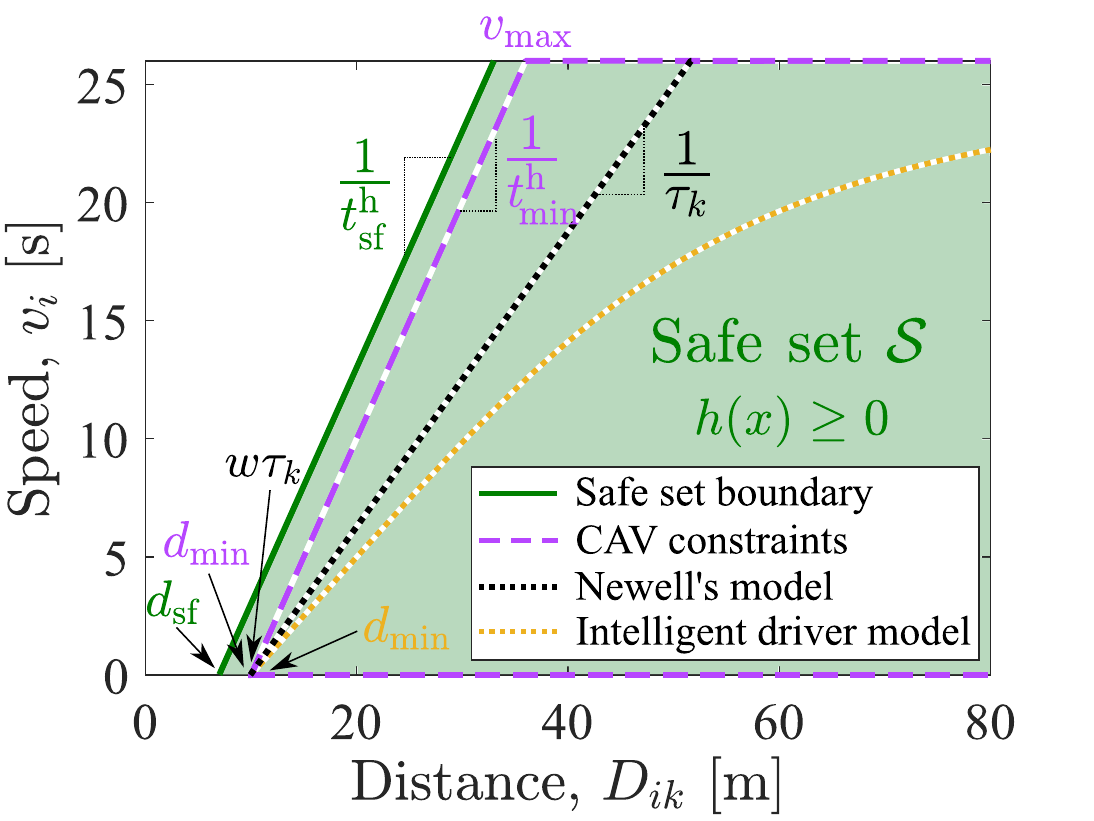}
  \end{center}
\caption{Safe set $\mathcal{S}$ (green region) visualized in state space for ${t^\mathrm{h}_\mathrm{sf}=1}$ $[\mathrm{s}]$ and ${d_\mathrm{sf}=7}$ $[\mathrm{m}]$. 
The range policies corresponding to the CAV constraints \eqref{eq:vconstraint} and \eqref{eq:rearend_constraint} (purple dashed), Newell's car following model \eqref{eq:newell} (black dotted) and the intelligent driver model \eqref{eq:u_IDM} (yellow dotted) are also indicated.} 
\label{fig:CBF}
\vspace{-5mm}
\end{figure}

To develop this safety filter, we consider a system involving a \CAV{i} and a preceding vehicle~$k$ traveling inside the control zone. 
Note that the preceding vehicle may be either an HDV or a CAV. 
Also, to ensure both rear-end safety and to avoid conflicts with vehicles on the other merging road, the preceding vehicle~$j$ refers to the vehicle either physically ahead of the \CAV{i} on the same lane, i.e., ${k = \max \, \{ j \in \SSS_i(t_i^0)\; | \; j < i \}}$,
or virtually projected from the neighboring lane, i.e., ${k = \max \, \{ j \in \NNN_i(t_i^0)\; | \; j < i \}}$, whichever is the closest in front. 
Based on \eqref{eq:model2}, the dynamics of such a system can be written as
\begin{equation}\label{eq:sys_CBF}
\begin{split}
\dot{D}_{ik}(t) &= v_{k}(t) - v_{i}(t), 
\\
\dot{v}_{i}(t) &= u_{i}(t), 
\end{split}    
\end{equation}
where ${D_{ik} := p_{k}-p_{i}}$ is the distance between vehicles $i$ and $k$. 
By defining the state ${x := [D_{ik}~v_{i}]^\top}$ and the input ${u := u_i}$, system \eqref{eq:sys_CBF} can be rewritten into the concise form
\begin{equation}\label{eq:sys_CBF_concise}
\dot{x} = f(x) + g(x)u,    
\end{equation}
with 
\begin{equation}\label{eq:fg}
f(x) = 
\begin{bmatrix}
v_{k} - v_{i}
\\
0
\end{bmatrix},
\quad
g(x) = 
\begin{bmatrix}
0
\\
1
\end{bmatrix}.
\end{equation}
We characterize the safety of \CAV{i} by the forward invariance of a  safe set ${\mathcal{S}\subset\mathbb{R}^2}$, that is, we require that
${x(0)\in \mathcal{S} \Rightarrow x(t)\in \mathcal{S}}$, $\forall t \ge 0$.
We define a scalar-valued control barrier function (CBF) such that ${h(x) \ge 0}$ when ${x \in \mathcal{S}}$, that is, safety can be ensured by maintaining the positivity of the CBF.
Here we define
\begin{align} \label{eq:S_h}
    \begin{split}
        \mathcal{S} &= \{x\in \mathbb{R}^2 : D_{ik} \ge d_\mathrm{sf} + t^\mathrm{h}_\mathrm{sf} v_{i}\}, 
        \\
        h(x) &= (D_{ik} - d_\mathrm{sf})/t^\mathrm{h}_\mathrm{sf} - v_{i},
    \end{split}
\end{align}
where ${d_\mathrm{sf} \ge 0}$ is the safe standstill distance while $t^\mathrm{h}_\mathrm{sf}$ is the safe time headway. 
An illustration of the safe set $\mathcal{S}$ is shown in Fig.~\ref{fig:CBF} where $t^\mathrm{h}_\mathrm{sf}$ and $d_\mathrm{sf}$ are indicated. 

It was shown in \cite{Tamas2023CDC} that since ${\nabla h(x) g(x) < 0}$ always holds, the positivity of the CBF defined in \eqref{eq:S_h} can be assured if the control action satisfies 
\begin{align}\label{eq:u_s_general}
   u \le u_\mathrm{s}(x) = - \frac{\nabla h(x)f(x)+\alpha(h(x))}{\nabla h(x)g(x)},
\end{align}
with ${\alpha>0}$. 
Substituting \eqref{eq:fg} and \eqref{eq:S_h} into \eqref{eq:u_s_general} we obtain
\begin{align}
    u_\mathrm{s}(x) = \frac{v_k - v_i}{t^\mathrm{h}_\mathrm{sf}} + \alpha\bigg(\frac{D_{ik}-d_\mathrm{sf}}{t^\mathrm{h}_\mathrm{sf}}-v_i\bigg).
\end{align}
Such safe control input can upper bound the CAVs' control inputs. 
In particular, denoting the optimal control input derived from Problems~\ref{prb:energy-optimal-1} and \ref{prb:optimal_MZ} as $u_\mathrm{o}$, a safety filter can be synthesized as
\begin{align}
    u = \min\{u_\mathrm{o},u_\mathrm{s}\},
\end{align}
which modifies the original optimal control input $u_\mathrm{o}$ to safe input $u_\mathrm{s}$ in a minimally invasive fashion. This ensures CAVs' conflict-free maneuvers under uncertain HDVs' behaviors.

\section{Simulation Results}\label{sec:sim}

In this section, we demonstrate the proposed framework by numerical simulations under various penetration rates of CAVs and various traffic volumes.
We consider a merging scenario for a control zone of length \SI{300}{m}, and we define a merging zone of length \SI{75}{m} upstream of the conflict point.
Vehicles enter the control zone with uniformly randomized initial speeds between $22$ and $\SI{26}{m/s}$.
We generate random vehicle entering time by a normal distribution where the traffic volumes determine the mean.

Given a preceding vehicle~$j$ (which may be projected from the neighboring road), the following \HDV{k}'s control input is given by
\begin{equation}\label{eq:u_IDM}
\!u_k \!= a \Bigg(\! 1 - \bigg(\! \frac{v_k}{v_{\max}} \!\bigg)^4 
\!-\! \Bigg(\! \frac{d_{\min} + t^\mathrm{h} v_k 
-\frac{v_k(v_j-v_k)}{2\sqrt{ab}}}{p_j-p_k} \!\Bigg)^{2} \Bigg),
\end{equation}
where $v_{\max}$ and $d_{\min}$ are the same as the speed limit and the standstill distance
imposed for CAVs in \eqref{eq:vconstraint} and
\eqref{eq:rearend_constraint} but here these denote desired values rather than constraints.
Also, $t^\mathrm{h}$ is the desired time headway, $a$ is the desired maximum acceleration, and $b$ is the comfortable deceleration. 
The parameters used in the simulations are 
${u_\mathrm{min} = \SI{-3}{m/s^2}}$, 
${u_\mathrm{max} = \SI{2}{m/s^2}}$, 
${v_\mathrm{max} = \SI{26}{m/s}}$, 
${t_{\min} = \SI{2}{s}}$,
${d_{\min} = \SI{10}{m}}$,
${t^\mathrm{h}_{\min} = \SI{1}{s}}$,
${d_\mathrm{sf} = \SI{7}{m}}$,
${t^\mathrm{h}_\mathrm{sf} = \SI{1}{s}}$,  ${\alpha = \SI{0.6}{1/s}}$,
${t^\mathrm{h} = \SI{2}{s}}$, 
${a = \SI{1}{m/s^2}}$, and
${b = \SI{1.5}{m/s^2}}$.

On the state space diagram in Fig.~\ref{fig:CBF}, we plot the range policy embedded in the IDM and compare it with Newell's car-following model (using the most aggressive $\tau_k$ value obtained from simulations). 
Both lie inside the safe set $\mathcal{S}$. 
We also remark that for simplicity of the simulation, when a feasible solution of Problems \ref{prb:energy-optimal-1} and \ref{prb:optimal_MZ} does not exist at the entrance of the control zone, the safe controller $u_\mathrm{s}$ is used for the CAV, instead of numerically solving a two-level optimization problem to derive $u_\mathrm{o}$.

\begin{figure}[!t]
  \begin{center}
  \includegraphics[scale=0.24]{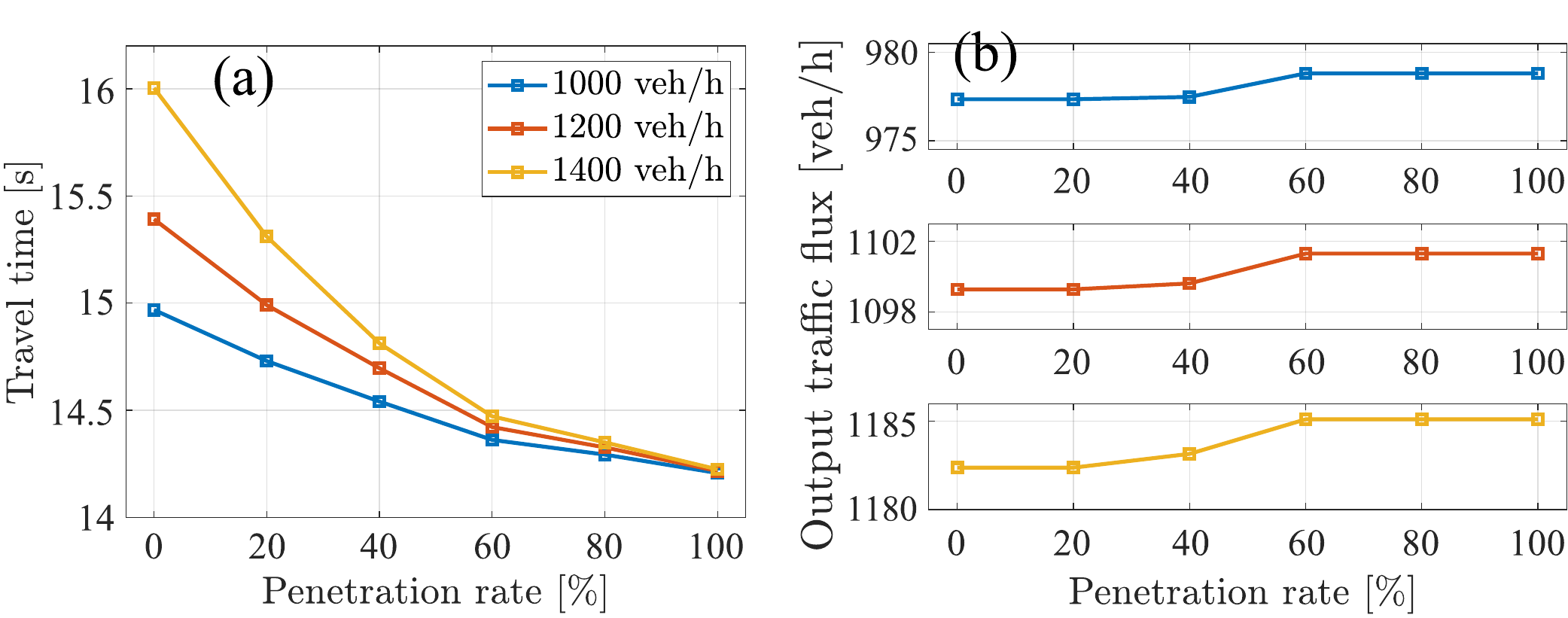}
  \end{center}

\caption{Average travel time (a) and output traffic flux (b) under different penetration rates of coordinated CAVs.}
\label{fig:statistical_results}
\vspace{-3mm}
\end{figure}

\begin{figure}[!t]
\begin{center}
  \includegraphics[scale=0.25]{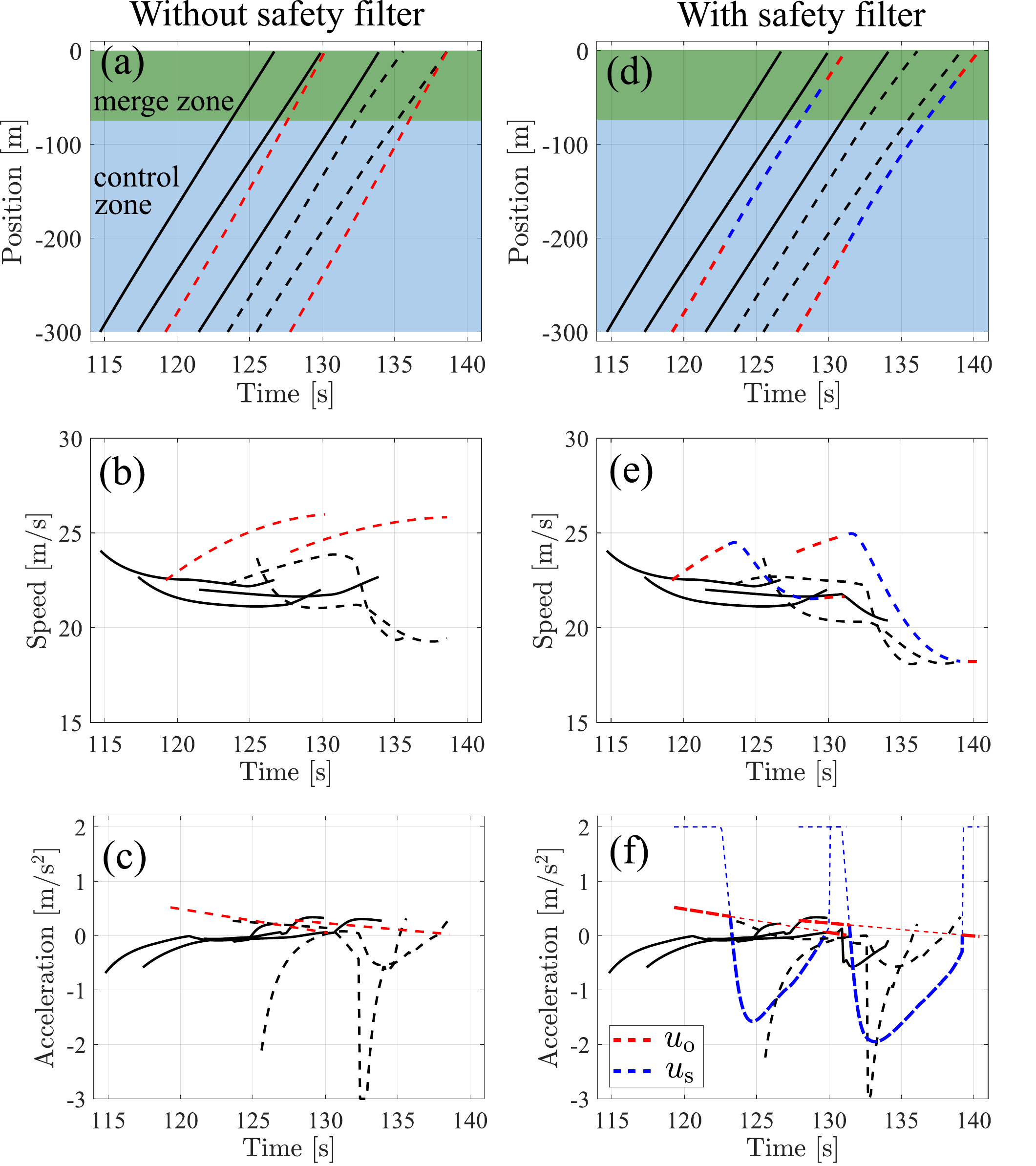}
\end{center}
\caption{Simulated position, speed, and acceleration profiles for a CAV and surrounding HDVs without and with CBF-based safety filter.
The trajectories of HDVs are represented by black curves. 
For CAV, red segments denote where optimal control input $u_\mathrm{o}$ is used while blue segments denote where safe input $u_\mathrm{s}$ is used. 
The actual control input used for the CAV is highlighted by the thick curve in panel (f).
The vehicles moving on different roads are distinguished by solid and dashed curves. The blue and green shaded areas highlight the control and merge zones.}
\vspace{-5mm}
\label{fig:CBF_simu}
\end{figure}

We conduct simulations using the proposed framework for six CAV penetration rates: 0\%, 20\%, 40\%, 60\%, 80\%, 100\%, and three traffic volumes: $1000$, $1200$, and $1400$ vehicles per hour. 
To quantify the benefits of coordinated CAVs, we use two metrics: average travel time in the control zone of the vehicles and the output traffic flux of the vehicles exiting the control zone. 
We conduct the simulations with 200 vehicles to compute these two metrics.
In Fig.~\ref{fig:statistical_results}, we summarize the results of average travel time and output flux for different CAV penetration rates and traffic volumes. 
Note that by increasing the CAV penetration rate, the average travel time improves significantly. 
Under high traffic volume (${\SI{1400}{veh/h}}$), $100 \%$ coordinated CAVs can improve the average travel time by $11 \%$ compared to the baseline traffic with pure HDVs.
Moreover, with higher CAV penetration, we also observe moderate improvements in the output traffic flux; see Fig.~\ref{fig:statistical_results}(b).

To demonstrate the CBF-based safety filter, the position, speed, and acceleration profiles of a few CAVs and HDVs are shown in Fig.~\ref{fig:CBF_simu} for traffic volume ${\SI{1200}{veh/h}}$.
The left panels reveal that without the safety filter, the optimal trajectories of the CAVs (red curves), derived at the entry of the control zone, may come very close to the trajectories of the HDVs (black curves). 
On the other hand, the right panels demonstrate that with the help of the safety filter, the CAVs can avoid conflicts with the HDVs. 
Note that in panels (d) and (e), the red segments represent where the optimal control input $u_\mathrm{o}$ is used for the CAV, while the blue segments represent where the safe input $u_\mathrm{s}$ is used. 
In panel (f), the control inputs $u_\mathrm{o}$ and $u_\mathrm{s}$ are plotted in red and blue, respectively, with the overall control input that the CAV used highlighted by the thicker curve.

\section{Concluding Remarks}\label{sec:conc}

In this paper, we presented a framework for coordinating CAVs in mixed traffic where they interact with HDVs.
We developed an upper-level optimization problem that yields the minimum travel time of the CAVs and uses the unconstrained trajectory solution of a low-level energy-optimal control problem.
We utilized Newell's car-following model with virtual projection to predict human driving behavior and introduced a control barrier function-based safety filter to address possible unsafe situations arising from the inaccuracy of these predictions.
Numerical simulations were used to demonstrate the performance and implications of the proposed framework.
Future research should address the problem of merging scenarios with multiple lanes and conduct experimental validation to assess the real-time practicality of the proposed framework.

\bibliographystyle{IEEEtran}
\bibliography{IEEEabrv,references,references_IDS}

\begin{thebibliography}{10}
\providecommand{\url}[1]{#1}
\csname url@samestyle\endcsname
\providecommand{\newblock}{\relax}
\providecommand{\bibinfo}[2]{#2}
\providecommand{\BIBentrySTDinterwordspacing}{\spaceskip=0pt\relax}
\providecommand{\BIBentryALTinterwordstretchfactor}{4}
\providecommand{\BIBentryALTinterwordspacing}{\spaceskip=\fontdimen2\font plus
\BIBentryALTinterwordstretchfactor\fontdimen3\font minus
  \fontdimen4\font\relax}
\providecommand{\BIBforeignlanguage}[2]{{%
\expandafter\ifx\csname l@#1\endcsname\relax
\typeout{** WARNING: IEEEtran.bst: No hyphenation pattern has been}%
\typeout{** loaded for the language `#1'. Using the pattern for}%
\typeout{** the default language instead.}%
\else
\language=\csname l@#1\endcsname
\fi
#2}}
\providecommand{\BIBdecl}{\relax}
\BIBdecl

\bibitem{ersal_connected_2020}
T.~Ersal, I.~Kolmanovsky, N.~Masoud, N.~Ozay, J.~Scruggs, R.~Vasudevan, and
  G.~Orosz, ``Connected and automated road vehicles: state of the art and
  future challenges,'' \emph{Vehicle System Dynamics}, vol.~58, no.~5, 2020.

\bibitem{Malikopoulos2020}
A.~A. Malikopoulos, L.~E. Beaver, and I.~V. Chremos, ``Optimal time trajectory
  and coordination for connected and automated vehicles,'' \emph{Automatica},
  vol. 125, no. 109469, 2021.

\bibitem{JinGabor2017}
J.~I. Ge and G.~Orosz, ``Optimal control of connected vehicle systems with
  communication delay and driver reaction time,'' \emph{IEEE Transactions on
  Intelligent Transportation Systems}, vol.~18, no.~8, pp. 2056--2070, 2017.

\bibitem{chalaki2020experimental}
B.~Chalaki, L.~E. Beaver, and A.~A. Malikopoulos, ``Experimental validation of
  a real-time optimal controller for coordination of cavs in a multi-lane
  roundabout,'' in \emph{IEEE Intelligent Vehicles Symposium (IV)}, 2020, pp.
  504--509.

\bibitem{katriniok2022fully}
A.~Katriniok, B.~Rosarius, and P.~M{\"a}h{\"o}nen, ``Fully distributed model
  predictive control of connected automated vehicles in intersections: Theory
  and vehicle experiments,'' \emph{IEEE Transactions on Intelligent
  Transportation Systems}, vol.~23, no.~10, pp. 18\,288--18\,300, 2022.

\bibitem{ChalakiCBF2022}
B.~Chalaki and A.~A. Malikopoulos, ``A barrier-certified optimal coordination
  framework for connected and automated vehicles,'' in \emph{Proceedings of the
  61th IEEE Conference on Decision and Control (CDC)}, 2022, pp. 2264--2269.

\bibitem{chalaki2020ICCA}
B.~Chalaki, L.~E. Beaver, B.~Remer, K.~Jang, E.~Vinitsky, A.~Bayen, and A.~A.
  Malikopoulos, ``Zero-shot autonomous vehicle policy transfer: From simulation
  to real-world via adversarial learning,'' in \emph{16th IEEE International
  Conference on Control \& Automation (ICCA)}, 2020, pp. 35--40.

\bibitem{chalaki2020hysteretic}
B.~Chalaki and A.~A. Malikopoulos, ``A hysteretic q-learning coordination
  framework for emerging mobility systems in smart cities,'' in \emph{2021
  European Control Conferences (ECC)}, 2021, pp. 17--22.

\bibitem{wang_social_2022}
``Social {Interactions} for {Autonomous} {Driving}: {A} {Review} and
  {Perspectives},'' \emph{Foundations and Trends{\textregistered} in Robotics},
  vol.~10, no. 3-4, pp. 198--376, 2022.

\bibitem{Yildiz2019}
B.~M. Albaba and Y.~Yildiz, ``Modeling cyber-physical human systems via an
  interplay between reinforcement learning and game theory,'' \emph{Annual
  Reviews in Control}, vol.~48, pp. 1--21, 2019.

\bibitem{Papageorigiou2020TRC}
M.~Karimi, C.~Roncoli, C.~Alecsandru, and M.~Papageorgiou, ``Cooperative
  merging control via trajectory optimization in mixed vehicular traffic,''
  \emph{Transportation Research Part C}, vol. 116, p. 102663, 2020.

\bibitem{Tomlin2019}
A.~Bajcsy, S.~L. Herbert, D.~{Fridovich-Keil}, J.~F. Fisac, S.~Deglurkar,
  A.~Dragan, and C.~Tomlin, ``A scalable framework for real-time multi-robot,
  multi-human collision avoidance,'' \emph{International Conference on Robotics
  and Automation (ICRA)}, pp. 936--943, 2019.

\bibitem{Althoff2021}
M.~Koschi and M.~Althoff, ``Set-based prediction of traffic participants
  considering occlusions and traffic rules,'' \emph{IEEE Transactions on
  Intelligent Vehicles}, vol.~6, no.~2, pp. 249--265, 2021.

\bibitem{Sanghoon2023ITSC}
S.~Oh, Q.~Chen, H.~E. Tseng, G.~Pandey, and G.~Orosz, ``Adaptive signalized
  intersection control in mixed traffic environment of connected vehicles with
  safety guarantees,'' in \emph{26th IEEE International Conference on
  Intelligent Transportation Systems (ITSC)}, 2023.

\bibitem{chalaki2021CSM}
B.~Chalaki, L.~E. Beaver, A.~M.~I. Mahbub, H.~Bang, and A.~A. Malikopoulos, ``A
  research and educational robotic testbed for real-time control of emerging
  mobility systems: From theory to scaled experiments,'' \emph{IEEE Control
  Systems}, vol.~42, no.~6, pp. 20--34, 2022.

\bibitem{buckman_sharing_2019}
N.~Buckman, A.~Pierson, W.~Schwarting, S.~Karaman, and D.~Rus, ``Sharing is
  caring: {Socially}-compliant autonomous intersection negotiation,'' in
  \emph{{IEEE}/{RSJ} {International} {Conference} on {Intelligent} {Robots} and
  {Systems} ({IROS})}.\hskip 1em plus 0.5em minus 0.4em\relax IEEE, 2019, pp.
  6136--6143.

\bibitem{liu2023safety}
H.~Liu, W.~Zhuang, G.~Yin, Z.~Li, and D.~Cao, ``Safety-critical and flexible
  cooperative on-ramp merging control of connected and automated vehicles in
  mixed traffic,'' \emph{IEEE Transactions on Intelligent Transportation
  Systems}, 2023.

\bibitem{faris_optimization-based_2022}
M.~Faris, P.~Falcone, and J.~Sj\"oberg, ``Optimization-based coordination of
  mixed traffic at unsignalized intersections based on platooning strategy,''
  in \emph{{IEEE} {Intelligent} {Vehicles} {Symposium} ({IV})}, 2022, pp.
  977--983.

\bibitem{newell_simplified_2002}
G.~F. Newell, ``A simplified car-following theory: a lower order model,''
  \emph{Transportation Research Part B}, vol.~36, no.~3, pp. 195--205, 2002.

\bibitem{CBF_2023}
A.~Alan, A.~J. Taylor, C.~R. He, A.~D. Ames, and G.~Orosz, ``Control barrier
  functions and input-to-state safety with application to automated vehicles,''
  \emph{IEEE Transactions on Control Systems Technology}, 2023,
  \url{https://doi.org/10.1109/TCST.2023.3286090}.

\bibitem{Tamas2023CDC}
T.~G. Moln\'ar, G.~Orosz, and A.~D. Ames, ``On the safety of connected cruise
  control: analysis and synthesis with control barrier functions,'' in
  \emph{62nd IEEE Conference on Decision and Control (CDC)}, 2023,
  \url{https://arxiv.org/abs/2309.00074}.

\bibitem{Minghao2023ITS}
M.~Shen, C.~R. He, T.~G. Molnar, A.~H. Bell, and G.~Orosz, ``Energy-efficient
  connected cruise control with lean penetration of connected vehicles,''
  \emph{IEEE Transactions on Intelligent Transportation Systems}, 2023,
  \url{https://doi.org/10.1109/TIV.2023.3281763}.

\bibitem{malikopoulos2018decentralized}
A.~A. Malikopoulos, C.~G. Cassandras, and Y.~J. Zhang, ``A decentralized
  energy-optimal control framework for connected automated vehicles at
  signal-free intersections,'' \emph{Automatica}, vol.~93, pp. 244--256, 2018.

\bibitem{molnar_board_2022}
T.~G. Moln\'ar, X.~A. Ji, S.~Oh, D.~Takács, M.~Hopka, D.~Upadhyay,
  M.~Van~Nieuwstadt, and G.~Orosz, ``On-board traffic prediction for connected
  vehicles: implementation and experiments on highways,'' in \emph{{American}
  {Control} {Conference} ({ACC})}, 2022, pp. 1036--1041.

\bibitem{VirtualPlatoon}
A.~I.~M. Medina, N.~van~de Wouw, and H.~Nijmeijer, ``Automation of a
  {T}-intersection using virtual platoons of cooperative autonomous vehicles,''
  in \emph{18th IEEE International Conference on Intelligent Transportation
  Systems (ITSC)}, 2015, pp. 1696--1701.

\bibitem{guo2019improved}
Y.~Guo, Q.~Sun, R.~Fu, and C.~Wang, ``Improved car-following strategy based on
  merging behavior prediction of adjacent vehicle from naturalistic driving
  data,'' \emph{IEEE Access}, vol.~7, pp. 44\,258--44\,268, 2019.

\end{thebibliography}

\balance

\end{document}